# Early Science Results from SOFIA, the World's Largest Airborne Observatory

James M. De Buizer
*Universities Space Research Association –
Stratospheric Observatory For Infrared Astronomy*

## ABSTRACT

The Stratospheric Observatory For Infrared Astronomy, or SOFIA, is the largest flying observatory ever built, consisting of a 2.7-meter diameter telescope embedded in a modified Boeing 747-SP aircraft. SOFIA is a joint project between NASA and the German Aerospace Center Deutsches Zentrum fur Luft und-Raumfahrt (DLR). By flying at altitudes up to 45000 feet, the observatory gets above 99.9% of the infrared-absorbing water vapor in the Earth's atmosphere. This opens up an almost uninterrupted wavelength range from 0.3-1600 microns that is in large part obscured from ground based observatories. Since its 'Initial Science Flight' in December 2010, SOFIA has flown several dozen science flights, and has observed a wide array of objects from Solar System bodies, to stellar nurseries, to distant galaxies. This paper reviews a few of the exciting new science results from these first flights which were made by three instruments: the mid-infrared camera FORCAST, the far-infrared heterodyne spectrometer GREAT, and the optical occultation photometer HIPO.

## 1. THE STRATOSPHERIC OBSERVATORY FOR INFRARED ASTRONOMY

SOFIA [1] is the world's largest flying observatory, consisting of a 2.7 m (diameter) reflective telescope that was developed by Deutsches Zentrum fur Luft und-Raumfart (DLR) that resides in a heavily modified Boeing 747-SP aircraft provided by NASA. The operation and development costs of SOFIA, as well as science and observing time, are divided up in 80:20 proportions split between NASA and DLR, respectively. The Universities Space Research Association (USRA) along with the Deutsches SOFIA Institut (DSI) in Germany is contracted by NASA and DLR, respectively, to oversee the science mission operations. These institutes are the main interface between SOFIA and the scientific community, and are responsible for operating SOFIA as a scientifically productive observatory.

The largest modification to the 747-SP aircraft was the insertion of the 18 ft (arc length) by 13.5 ft-wide opening in fuselage, that is covered by a door system, which when opened allows the 40000-pound infrared telescope behind it to peer into the heavens. Fig. 1a shows SOFIA during a test flight with its cavity door opened. Fig. 1b shows a cut-away view of the aircraft. The telescope itself is located aft of a reinforced pressure bulkhead, and when the cavity door is open, everything in this part of the aircraft (including the telescope) is exposed to the low temperatures and pressures that occur at high altitude. Everything forward of the pressure bulkhead is pressurized and climate controlled in a similar fashion to a commercial airliner. It is in this main cabin that the scientists and mission control specialists command the telescope and instruments and conduct their scientific studies. Because of its accessibility and ability to carry passengers, SOFIA has a vigorous and highly visible Education and Public Outreach program, and thus there are also areas designated for educators on the flights as well.

The main reason for developing an airborne platform for infrared astronomy studies is that the Earth's atmosphere contains water vapor which readily absorbs infrared photons from space before they reach the ground. By flying at altitudes up to 45000 ft, SOFIA observes from above more than 99% of the Earth's atmospheric water vapor. Fig. 1c shows the typical atmospheric transmission observed at wavelengths 1 to 1000 µm as seen from SOFIA compared to one of the best ground-based observing locations on the Earth, the summit of Mauna Kea in Hawaii. This figure shows that SOFIA opens up observing windows to the universe not available from the ground, including the expansive 25 to 350 µm wavelength regime.

Of course another way of getting above Earth's atmosphere is to launch an observatory into space. While space observatories have many merits, the airborne platform of SOFIA also has its share of advantages. Most importantly, SOFIA can be considered a near-space observatory that comes home after every flight. This allows for its science instruments to be exchanged regularly for repairs and upgrades without the need for costly manned space missions. It can accommodate rapidly changing science requirements and incorporate new technologies quickly. It can also serve as a test bed for space technologies that will eventually be launched on satellites without the need for the instruments to be space qualified first.

SOFIA will offer international science teams approximately 1000 high-altitude science observing hours per year during its two-decade designed lifetime. A suite of first generation instruments have already been chosen to cover most of the operating wavelengths of SOFIA with a variety of spectral ranges as shown in Fig. 1d. The three of these instruments (HIPO, FORCAST, and GREAT) have already produced science, and one science highlight for each of these instruments will be presented in the rest of this article.

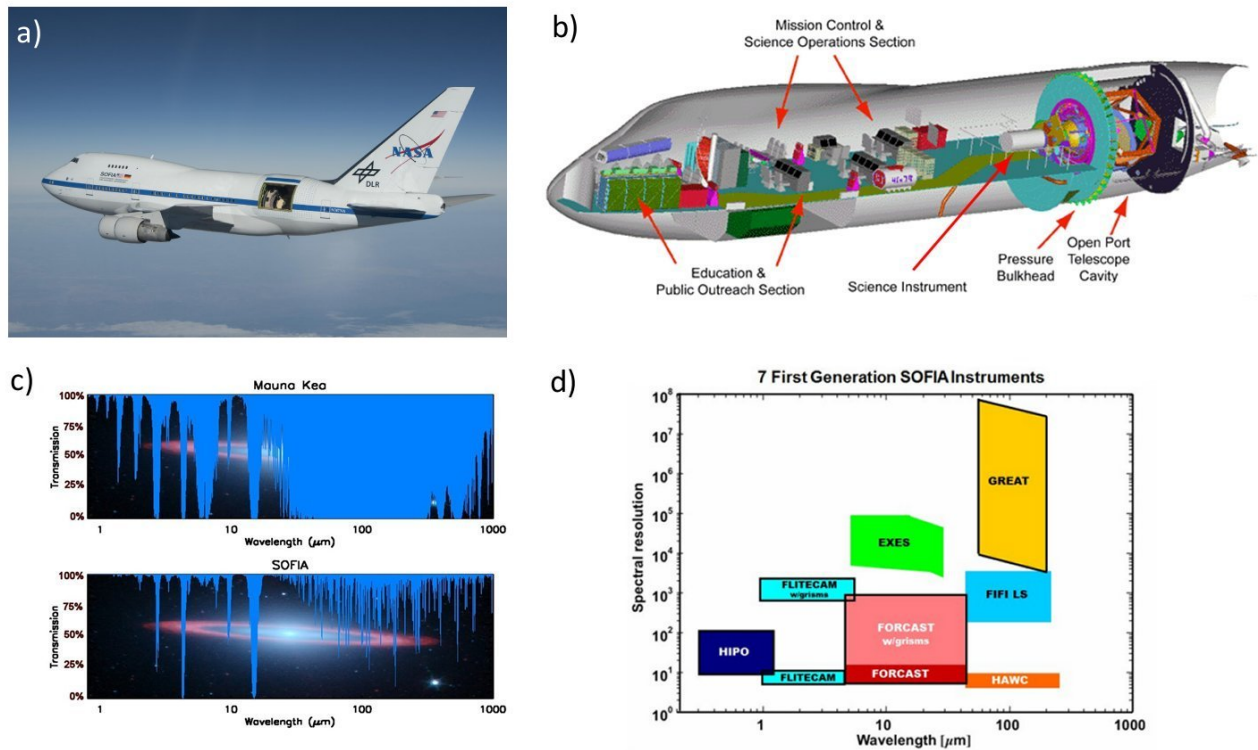

Fig. 1. a) The SOFIA aircraft with its door open during a daytime test flight (Image credit: NASA/Jim Ross); b) A cut-away view of the fuselage of SOFIA with the main areas of interest labeled; c) The typical atmospheric transmission of infrared light as a function of wavelength as seen by SOFIA and from the observatories on the summit of Mauna Kea, Hawaii; d) A plot of the infrared wavelengths covered by SOFIA's suite of first generation instruments, and their spectral resolution coverages.

## 2. PROBING THE ATMOSPHERE OF PLUTO WITH HIPO

The dwarf planet Pluto has a strange orbit that actually crosses over the orbit of Neptune. In fact, from 1979 to 1999 Neptune was farther from the Sun than Pluto (Fig. 2a). When Pluto is at the point farthest from the Sun in its orbit it is almost twice the distance away as when it is at its closest point. Consequently, Pluto experiences quite a large range of temperatures during a full orbit of the Sun. Pluto is known to have an atmosphere, however as it makes its way along its orbit farther and farther from the Sun, it is uncertain if and when the gases in Pluto's atmosphere will freeze and precipitate to the dwarf planet's surface.

HIPO (High-speed Imaging Photometer for Occultation) is a special-purpose science instrument for SOFIA that is designed to provide simultaneous high-speed time resolved imaging at two optical wavelengths [2]. The main scientific goal of HIPO is the observation of stellar occultations. In a stellar occultation, a star serves as a probe of the atmospheric structure of a solar system object or the surface density structure of a planetary ring or comet. When a solar system object passes in front of a star, it dims the light from the star. As can be seen in Fig. 2b, plotting this light intensity versus time yields a differently shaped light curve for an object with or without an atmosphere. It was actually this technique that was used by SOFIA's predecessor, the Kuiper Airborne Observatory (KAO), to discover the atmosphere of Pluto in 1988 when Pluto was near its closest point to the Sun [3]. This same technique was used by HIPO to measure the light curve of Pluto in June of last year as the dwarf planet passed in front of a star, showing that the atmosphere still exists (Fig. 2c). Images taken before, during, and after the occultation in Fig. 2d show the dimming of the star during the occulation as Pluto passed in front of it and blocked its light [4].

Such observations are very difficult to perform without the mobility of SOFIA (or, in the past, KAO). Like a solar eclipse, an occultation has a narrow "path of totality" across the Earth's surface where one needs to be to observe the shadow of the occultation pass over. The chances of this path going over an established observatory are small. However, due to the mobility of SOFIA, the aircraft was able to intercept the shadow of the occulation 1800 miles off the west coast of the United States as it raced across the Pacific Ocean at more than 53,000 mph.

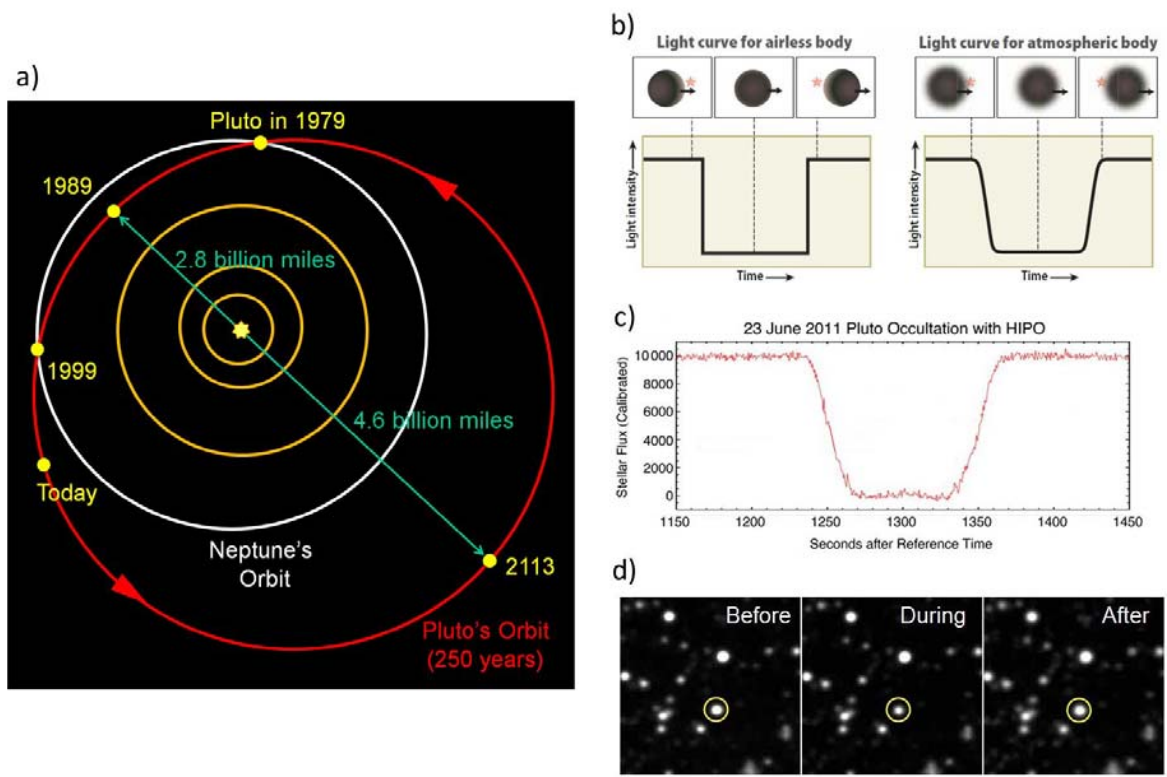

Fig. 2. a) A schematic diagram of the orbits of Pluto (red) and Neptune (white). Yellow orbits are for Uranus, Saturn, and Jupiter (the terrestrial planet orbits are too small to show in this diagram); b) Two plots showing the difference between a stellar occultation of a body with and without an atmosphere (Image credit: Astronomy Magazine/Roen Kelly); c) The observed light curve of the occultation of Pluto in June of 2011 taken by HIPO on SOFIA showing the rounded edges and sloping angles typical of an object with an atmosphere (Image credit: Ted Dunham/Lowell Observatory); d) Optical images taken with the guide camera of SOFIA showing the stellar occultation just before, during, and just after Pluto crossed in front of the star. The star and Pluto are too close in each frame to see them separately but the dimming is noticeable in the middle frame (Image credit: DLR/U. Stuttgart/DSI/J. Wolf/E. Pfuller/M. Wiedemann).

## 3. THE DISCOVERY OF THE MOLECULE SH IN SPACE WITH GREAT

Diatomic hydrides (molecules containing one atom of hydrogen, H, and one additional atom) represent the simplest of molecules and trace key astrophysical processes in the interstellar environment. After the discovery of the first such molecule in space, CH, by [5] in 1937, only 6 further diatomic hydrides have ever been discovered. With the exception of $H_2$ (i.e., "HH") discovered by [6], all of the diatomic hydrides discovered have come from the same part of the periodic table – the non-metals (Fig. 3a). The glaring exceptions are sulfur and potassium, which occupy the same area of the periodic table, and hence should also just as readily bond with H atoms.

In its ground rotational state, SH can absorb radiation in a lambda doublet near 1.383 THz ($\lambda \approx 217$ μm), a frequency completely inaccessible from the ground and one that falls between the gaps in coverage by Herschel Space Observatory. However, SOFIA's GREAT (German REceiver for Astronomy at Terahertz frequencies) spectrometer covers this frequency with its 63-250 μm wavelength range (i.e., 4.8-1.2 THz frequency range) with high spectral resolution [7].

To maximize the chance of detection, SOFIA observed the very bright far-IR star forming region W49N, which lies almost halfway across the Galaxy from the Sun (Fig. 3b). Any SH gas that exists in the Galaxy between us and W49N should be readily seen in absorption against the bright emission from W49N. Indeed, SOFIA and GREAT did in fact detect the spectral signature of the lambda doublet of SH (Fig. 3c) for the first time [8].

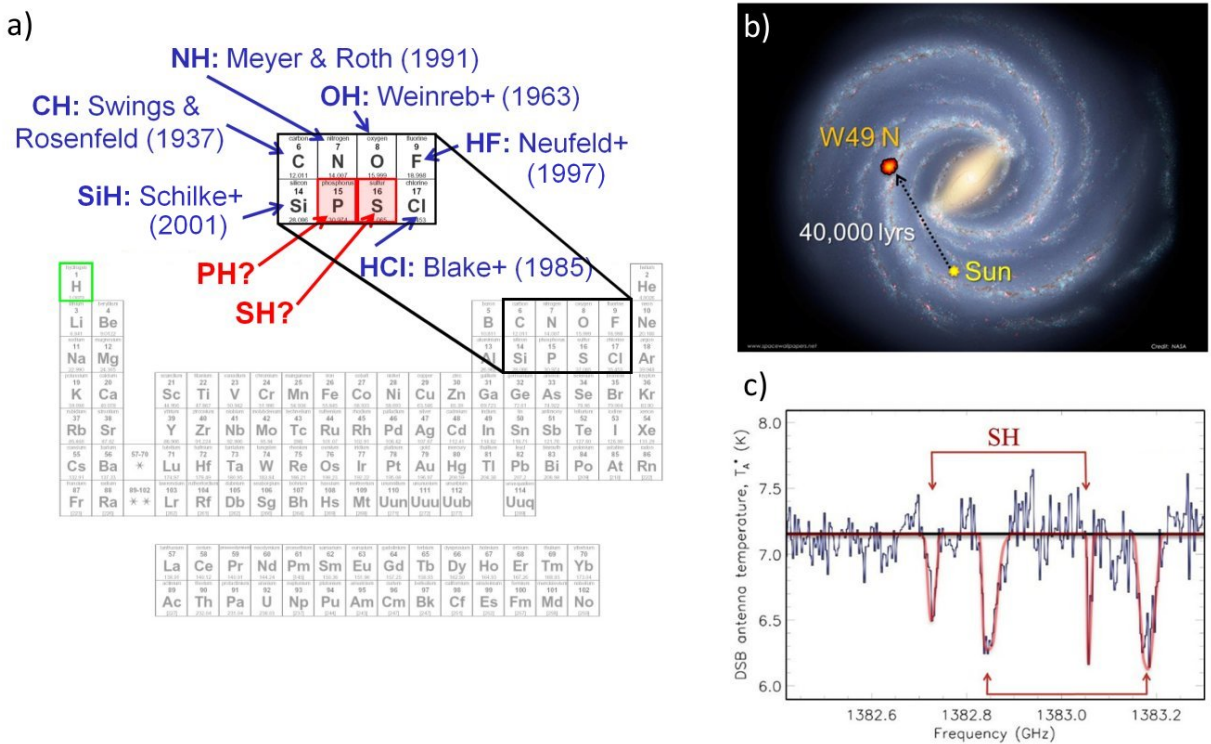

Fig. 3. a) A view of the periodic table with a zoom-in on a portion containing the non-metals. The citations for the discoveries of several diatomic hydrides are shown ([5],[9],[10],[11],[12],[13]), with SH and PH being the exceptions; b) A cartoon view of the Milky Way from above showing the locations of the Sun and the extremely bright and distance star-forming region W49N (Background image credit: NASA); c) A spectrum taken toward W49N showing the discovery of the lambda doublet of SH in absorption from [8].

## 4. PEERING INTO THE HEART OF THE ORION NEBULA WITH FORCAST

The Orion nebula is the closest massive star-forming region to the Sun, lying nearly 1400 lightyears away from us [14]. At the center of the Orion nebula lies the extremely bright and active areas of star formation known as the BN/KL region and the Ney-Allen region (Fig. 4a). A long-standing mystery for astronomers that study this complex region is determining what objects(s) provide the bulk of the energy heating the BN/KL region. This question has been difficult to answer because the region is shrouded in a cocoon of dense dust and gas, shielding it from view in the visible. However, light from sources inside at wavelengths greater than ~20 μm can penetrate through this dust allowing us to see their emission. Furthermore, there are many objects very close to each other in the BN/KL region, requiring a large telescope to resolve the infrared light into its separate emitting objects. Consequently, SOFIA proved to be perfectly suited to investigate this region in detail.

This nebula was the subject of investigation by SOFIA's "first light" instrument, FORCAST (Faint Object infra-Red CAmera for the SOFIA Telescope). FORCAST is a 5-40 μm imager that is also capable of low-to-medium resolution (R=100-1500) mid-infrared spectroscopy [15]. FORCAST images taken at multiple wavelengths yielded the fluxes of several sources in the BN/KL region from 6-38 μm (Fig. 4a).

There are multiple sources of infrared emission previously identified in this region (i.e., [16]). It is believed that only a few are massive (proto)stars, and that the rest are merely clumps of gas and dust that are externally heated by those massive (proto)stars. The source designated "BN" in the inset in Fig. 4a is believed to be one such massive (proto)star responsible for at least some of the heating of the surrounding BN/KL region [17].

The fluxes of each source seen by FORCAST were plotted as a function of wavelength and attempts were made to try to fit the flux distributions with massive star formation models to determine the true nature of each source. Fig. 4b shows these plots for two such sources, BN and IRc4. Thanks to the FORCAST data, it was concluded for the first time that sources BN and IRc4 are not only embedded massive stars, but have such large luminosities that they are responsible for a significant part of the heating in this part of the Orion nebula [18]. This result was surprising, as previous studies had misidentified IRc4 as just a passive clump of dust and gas [i.e., 16].

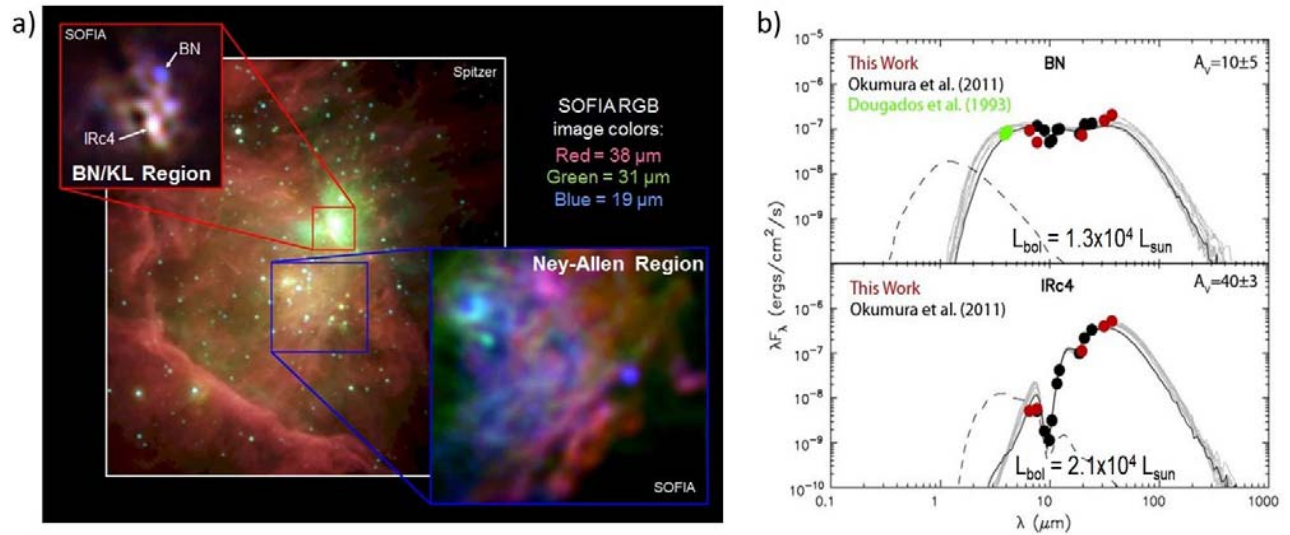

Fig. 4. a) A background image from the Spitzer Space telescope shows the central region of the Orion Nebula in the near-infrared. The BN/KL region is contained in the red box, while the Ney-Allen Region is contained in the blue box. The BN/KL region is saturated and sources within are unresolved in the Spitzer image. The SOFIA image of the BN/KL region (inset red box), is comprised of longer wavelengths of emission and with high enough resolution to separate out individual sources; b) For two sources, BN and IRc4, plots of their fluxes as a function of wavelength are shown. The red data points come from the SOFIA observations, the others are from the literature. Grey lines show the best fits from massive star formation models, and the median bolometric luminosity given by those fits is given for each source. These luminosities are so high that they must be massive proto(stars). Image from [18].

## 5. CONCLUSIONS

In the last couple of years, SOFIA has transitioned from a project in development into a full-fledged astronomical observatory. It is now producing exciting world-class science, only a few of which have been highlighted in this article. More information of the results presented here along with much more science from SOFIA's first 30 flights with FORCAST[1] and GREAT[2] have been presented in two dedicated peer-reviewed volumes. With the range of current and future instrumentation, and with the projected 20 year operational lifetime, it is expected that SOFIA will yield a vast and rich collection of scientific results and be a premier infrared facility for years to come.

---

[1] The FORCAST special issue contained eight articles and was published in the 2012 April 20 issue of the Astrophysical Journal Letter (Volume 749, Number 2).
[2] The GREAT special issue contained sixteen articles and was published in the 2012 June issue of the journal of Astronomy & Astrophysics (Volume 542).